\documentclass[a4paper,12pt]{spieman}  
\usepackage{amsmath,amsfonts,amssymb}
\usepackage{graphicx}
\usepackage{setspace}
\usepackage{pdfpages}
\usepackage{tocloft}
\usepackage[printonlyused,nolist]{acronym}
\usepackage{xcolor}
\definecolor{dkfzblue}{rgb}{0.349,0.529,0.824}
\usepackage[skins,breakable]{tcolorbox}

\begin{acronym}[MMMMMMMM]
\acro{2D}{2-Dimensional}
\acro{3D}{3-Dimensional}
\acro{4D}{4-Dimensional}
\acro{CC}{Correlation Coefficient}
\acro{CNN}{Convolutional Neural Network}
\acro{ICG}{Indocyanine Green}
\acro{MAE}{Mean Absolute Error}
\acro{MRE}{Mean Relative Error}
\acro{MSE}{Mean Squared Error}
\acro{PA}{Photoacoustic}
\acro{PAI}{Photoacoustic Imaging}
\acro{PSNR}{Peak Signal-to-Noise Ratio}
\acro{SNR}{Signal-to-Noise Ratio}
\acro{SSIM}{Structural Similarity}
\end{acronym}

\title{Deep learning for biomedical photoacoustic imaging: A review}

\author[a,b,*]{Janek Gr{\"o}hl}
\author[a]{Melanie Schellenberg}
\author[a,c]{Kris Dreher}
\author[a,b,d,*]{Lena Maier-Hein}
\affil[a]{German Cancer Research Center, Computer Assisted Medical Interventions, Heidelberg, Germany}
\affil[b]{Heidelberg University, Medical Faculty, Heidelberg, Germany}
\affil[c]{Heidelberg University, Faculty of Physics and Astronomy, Heidelberg, Germany}
\affil[d]{Heidelberg University, Faculty of Mathematics and Computer Science, Heidelberg, Germany}

\cftpagenumbersoff{figure}
\cftpagenumbersoff{table} 
\begin{document} 
\maketitle

\begin{abstract}
\ac{PAI} is a promising emerging imaging modality that enables spatially resolved imaging of optical tissue properties up to several centimeters deep in tissue, creating the potential for numerous exciting clinical applications. However, extraction of relevant tissue parameters from the raw data requires the solving of inverse image reconstruction problems, which have proven extremely difficult to solve. The application of deep learning methods has recently exploded in popularity, leading to impressive successes in the context of medical imaging and also finding first use in the field of \ac{PAI}. Deep learning methods possess unique advantages that can facilitate the clinical translation of \ac{PAI}, such as extremely fast computation times and the fact that they can be adapted to any given problem. In this review, we examine the current state of the art regarding deep learning in \ac{PAI} and identify potential directions of research that will help to reach the goal of clinical applicability.\\
\end{abstract}


{\noindent \footnotesize\textbf{*}Corresponding authors: JG  \linkable{j.groehl@dkfz-heidelberg.de} and LMH \linkable{l.maier-hein@dkfz-heidelberg.de}}

\section{Introduction}
\label{sec:intro}

\acf{PAI} is a comparatively young and rapidly emerging imaging modality that promises real-time, noninvasive, and radiation-free measurement of optical tissue properties\cite{wang2014photoacoustic}. In contrast to other optical imaging modalities, \ac{PAI} induces the emergence of acoustic signals to enable structural imaging of chromophores - molecular structures that absorb light - up to several centimeters deep into the tissue. This high depth penetration is possible because the acoustic scattering of the arising sound waves is orders of magnitude smaller than the optical scattering of the incident light in biological tissue. The underlying physical principle for signal generation is the \emph{PA effect}\cite{rosencwaig1976theory}. It is induced by extremely short light pulses that cause an initial pressure rise $p_0$ inside the tissue. The initial pressure $p_0 = \Gamma \cdot \mu_a \cdot \phi$ is proportional to the optical absorption coefficient $\mu_a$, the local light fluence $\phi$, and the temperature-dependent Gr\"uneisen parameter $\Gamma$. The deposited energy is released in form of sound waves that can be measured as time-series pressure data $p(t) = A(p_0, \theta)$ with appropriate acoustic detectors, such as ultrasound transducers. Here, acoustic forward operator $A$ operates on the initial pressure distribution $p_0$ taking into consideration the acoustic properties $\theta$ of the medium.\\

Due to its rapid development, \ac{PAI} has seen various clinical application attempts over the last few years. Among these, cancer research is a field where \ac{PAI} shows serious potential \cite{mallidi2011photoacoustic,li2019benign,zhang2018photoacoustic,quiros2018optoacoustics,oh2006three,weight2006photoacoustic,zhang2010subwavelength,zhang2010chronic,song2008noninvasive,erpelding2010sentinel,garcia2015dual}. In this use case, hemoglobin is the enabling endogenous chromophore, due to amplified and sustained angiogenesis \cite{hanahan2011hallmarks} being one of the hallmarks of cancer and due to the cancer cells' increased metabolism, which potentially induces a decrease in local blood oxygenation \cite{wang2012biomedical}. Furthermore, because inflammatory processes also change the hemodynamic behavior of tissue, \ac{PAI} is also used for imaging of inflamed joints \cite{wang2007noninvasive,rajian2012photoacoustic,jo2018detecting} or staging of patients with Crohn's disease \cite{knieling2017multispectral,waldner2016multispectral,lei2019characterizing}. To further increase the potential of \ac{PAI}, it is also applied in combination with other imaging modalities, especially ultrasound imaging \cite{niederhauser2005combined,aguirre2011potential,needles2013development,garcia2015dual,elbau2017quantitative,mandal2019multimodal}. \ac{PAI} is further used for brain imaging \cite{wang2003noninvasive,ku2005imaging,hu2009functional,yao2014photoacoustic,mohammadi2019skull} or surgical and interventional imaging applications, such as needle tracking \cite{kim2010handheld,su2010photoacoustic}.\\

The signal contrast of \ac{PAI} is caused by distinct wavelength-dependent absorption characteristics of the chromophores\cite{upputuri2016recent}. But to exploit information of $\mu_a$ for answering clinical questions, open research questions remain that can be categorized into four main areas. In the following, we explain these four major categories and summarize their principal ideas.\\

\textbf{Acoustic inverse problem.} The most pressing problem concerns the reconstruction of an image $S$ from recorded time-series data by estimating the initial pressure distribution $p_0$ from $p(t)$. This problem is referred to as the acoustic inverse problem. To this end, an inverse function $A^{-1}$ for the acoustic operator $A$ needs to be computed in order to reconstruct a signal image $S = A^{-1}(p(t)) \approx p_0 = \mu_a \cdot \phi \cdot \Gamma$ that is an approximation of $p_0$. Typical examples of algorithms to solve this problem are the universal back-projection\cite{xu2005universal}, delay-and-sum \cite{mozaffarzadeh2017double}, time reversal\cite{treeby2010k}, or iterative reconstruction schemes\cite{huang2013full}. While the acoustic inverse problem can be well-posed in certain scenarios (for example by using specific detection geometries) and thus can have a unique solution, several factors lead to considerable difficulties in solving it. These include wrong model assumptions\cite{cox2009challenges}, limited-view\cite{waibel_reconstruction_2018} and limited-bandwidth detectors\cite{buchmann2017characterization}, or device modeling\cite{sahlstrom2020modeling} and calibration errors\cite{buchmann2020quantitative}.\\

\textbf{Image post-processing.} \ac{PAI}, in theory, has exceptionally high contrast and spatial resolution\cite{xu2006photoacoustic}. Because the acoustic inverse problem is ill-posed in certain settings and because of the presence of noise, many reconstructed PA images suffer from distinct artifacts. This can cause the actual image quality of a PA image to fall short of its theoretical potential. To tackle these problems, image post-processing algorithms are being developed to mitigate the effects of artifacts and noise and thus improve overall image quality.\\

\textbf{Optical inverse problem.} Assuming that a sufficiently accurate reconstruction of $p_0$ from $p(t)$ has been achieved, the second principle problem that arises is the estimation of the underlying optical properties (most importantly the absorption coefficient $\mu_a$). It is an inverse problem and is referred to as the optical inverse problem. Furthermore, the problem has proven to be exceptionally involved, which can be derived by the fact that methods to solve the problem have not been successfully applied to \emph{in vivo} data yet. It belongs to the category of ill-posed inverse problems, as it does not necessarily possess a unique solution. Furthermore, several other factors make it hard to tackle, including wrong model assumptions\cite{cox2009challenges}, non-uniqueness and non-linearity of the problem\cite{shao2011estimating}, spectral coloring\cite{tzoumas2016eigenspectra}, and the presence of noise and artifacts\cite{kazakeviciute2016multispectral}. Quantification of the absorption coefficient has, for example, been attempted with iterative reconstruction approaches\cite{cox2006two}, via fluence estimation\cite{brochu2016towards}, or by using machine learning-based approaches\cite{kirchner2018context}.\\

\textbf{Semantic image annotation.} Based on the diagnostic power of optical absorption it is possible to generate semantic image annotations of multispectral PA images and a multitude of methods for it are being developed to specifically tackle questions of clinical relevance. To this end, algorithms are being developed that are able to classify and segment multispectral PA images into different tissue types and that can estimate clinically relevant parameters that are  indicative of a patient's health status (such as blood oxygenation). Current approaches to create such semantic image annotations suffer from various shortcomings, such as long computation times or the lack of reproducibility in terms of accuracy and precision when being applied to different scenarios.\\

Simultaneously to the rapid developments in the field of \ac{PAI}, deep learning algorithms have become the \emph{de facto} state of the art in many areas of research\cite{shen2017deep} including medical image analysis. A substantial variety of medical applications include classical deep learning tasks such as disease detection \cite{liu2019comparison}, image segmentation \cite{tajbakhsh2020embracing}, and classification \cite{tandel2019review}. Recently, deep learning has also found entrance into the field of \ac{PAI}, as it promises unique advantages to solve the four listed problems, thus promoting clinical applicability of the developed methods. One further prominent advantage of deep learning is the extremely fast inference time, which enables real-time processing of measurement data.\\

This review paper summarizes the development of deep learning in \ac{PAI} from the emergence of the first PA applications in 2017 until today and evaluates progress in the field based on the defined task categories. In section \ref{sec:methods}, we outline the methods for our structured literature research. General findings of the literature review including the topical foci, data acquisition techniques, used simulation frameworks as well as network architectures are presented in section \ref{sec:general}. The reviewed literature is summarized according to the four principal categories in sections \ref{sec:reconstruction} to \ref{sec:clinical_translation}. Finally, the findings are discussed and summarized in section \ref{sec:discussion}.\\

\section{Methods of literature research}
\label{sec:methods}

\begin{figure}[h!tb]
    \centering
    \includegraphics{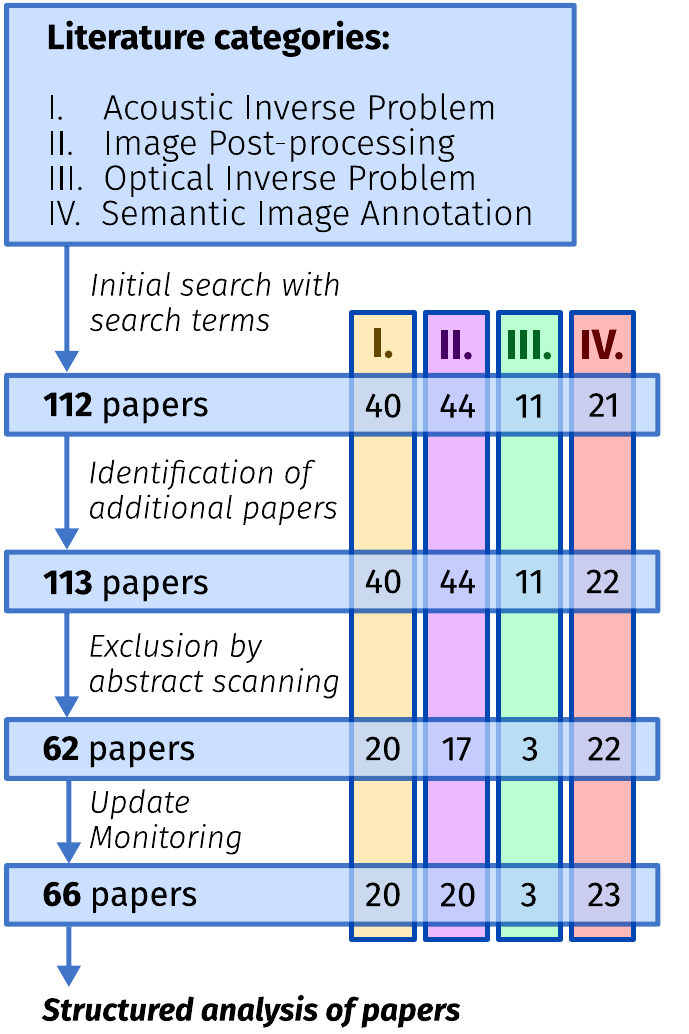}
    \caption{Overview of the literature review algorithm. First, potentially fitting papers are identified based on an initial search. The search results are complemented by adding additional papers found by other means than the search engines, and finally, non-relevant papers are excluded by removing duplicates and by abstract scanning.}
    \label{fig:search_algorithm}
\end{figure}

Above, we described the \ac{PAI}-specific challenges that deep learning can be applied to and thus divided the topic into four major categories: \emph{I. Acoustic inverse problem}, \emph{II. Image post-processing}, \emph{III. Optical inverse problem}, and \emph{IV. Semantic image annotation}. We conducted a systematic literature review for the period between January 2017 and September 2020 and assigned each identified paper to the most suitable categories. For the search, we used several scientific search engines: \emph{Google Scholar}, \emph{IEEE Xplore}, \emph{Pubmed}, \emph{Microsoft Academic Search Engine}, and the \emph{arXiv} search function with the search string \textbf{("Deep learning” OR "Neural Network") AND ("photoacoustic" OR "optoacoustic")}. The search results were then refined in a multi-step process (see Fig. \ref{fig:search_algorithm}).\\

First, potential candidates were identified based on an initial search using their title, as well as the overview presented by the search engine. The search results were complemented by adding additional papers found by means other than the named search engines. For this purpose, we specifically examined proceedings of relevant conferences, websites of key authors we identified, and websites of PA device vendors. Finally, non-relevant papers were excluded by removing Journal/Proceeding paper duplicates and by abstract scanning to determine whether the found papers match the scope of this review. Using the abstract, we excluded papers that did not apply deep learning, and those that did not match the scope of \emph{biomedical} \ac{PAI}. The remaining articles were systematically examined by the authors using a questionnaire to standardize the information that was to be extracted. While writing the paper, we continuously monitored the mentioned resources for new entries until the end of September 2020.\\

In total, applying the search algorithm as detailed above, we identified 66 relevant papers (excluding duplicates and related, but out-of-scope work) that have been published since 2017.

\section{General findings}
\label{sec:general}

The application of deep learning techniques to the field of \ac{PAI} has constantly been accelerating over the last three years and has simultaneously generated a noticeable impact on the field.\\

\begin{figure}[h!tb]
    \centering
    \includegraphics{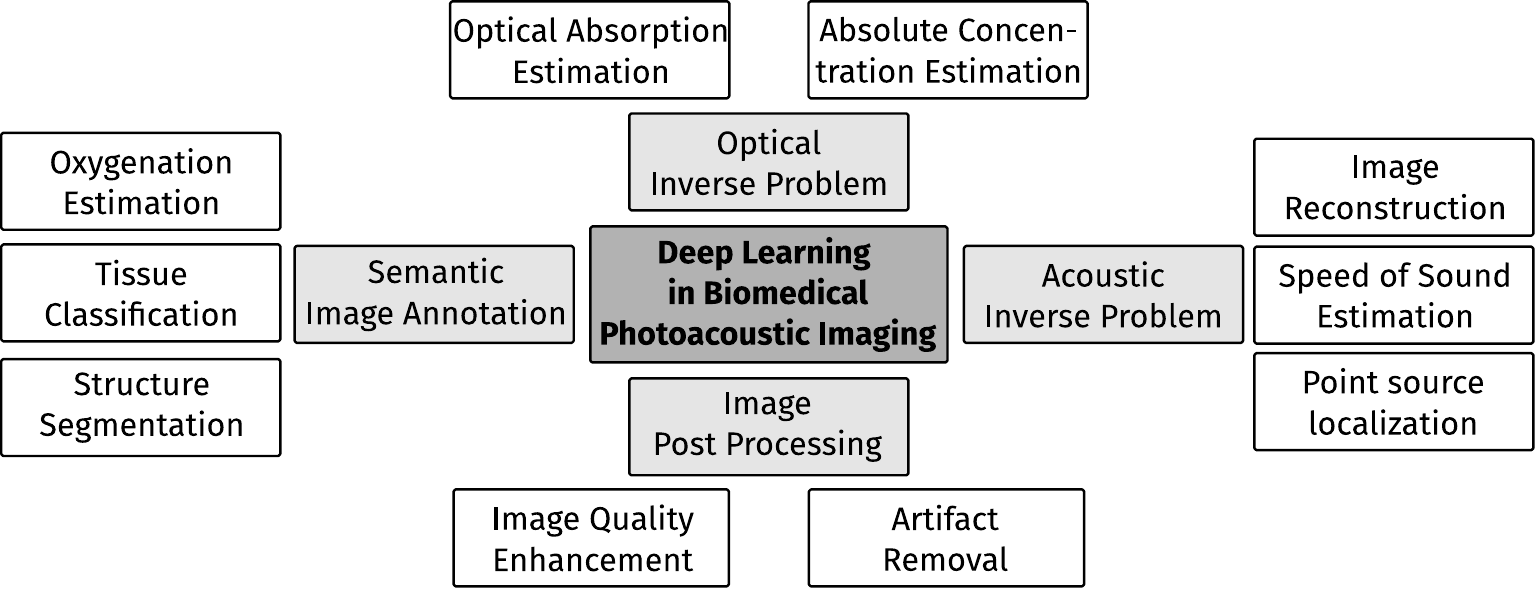}
    \caption{Overview over the topical foci of current research towards applying deep learning algorithms to problems in biomedical \ac{PAI}.}
    \label{fig:topics}
\end{figure}

\textbf{Topical foci.} After categorization of the papers into the predetermined four areas, the papers were arranged into thematic subcategories (see Fig. \ref{fig:topics}). Papers related to the {Acoustic Inverse Problem} (section \ref{sec:reconstruction}) generally focus on the reconstruction of PA images from raw time-series pressure data but also related topics, such as dealing with limited-view or limited data settings, as well as the estimation of the speed of sound of tissue. The {Image Post-processing} (section \ref{sec:processing}) category entails papers that deal with image processing algorithms that are being applied after image reconstruction. The aim of such techniques usually is to improve the image quality, for example by noise reduction or artifact removal. The three papers assigned to the {Optical Inverse Problem} (section \ref{sec:quantification}) deal with the estimation of absolute chromophore concentrations from PA measurements. Finally, papers dealing with Semantic Image Annotation (section \ref{sec:clinical_translation}) tackle use cases, such as the segmentation and classification of tissue types or the estimation of functional tissue parameters, such as blood oxygenation.\\

\begin{figure}[h!tb]
    \centering
    \includegraphics{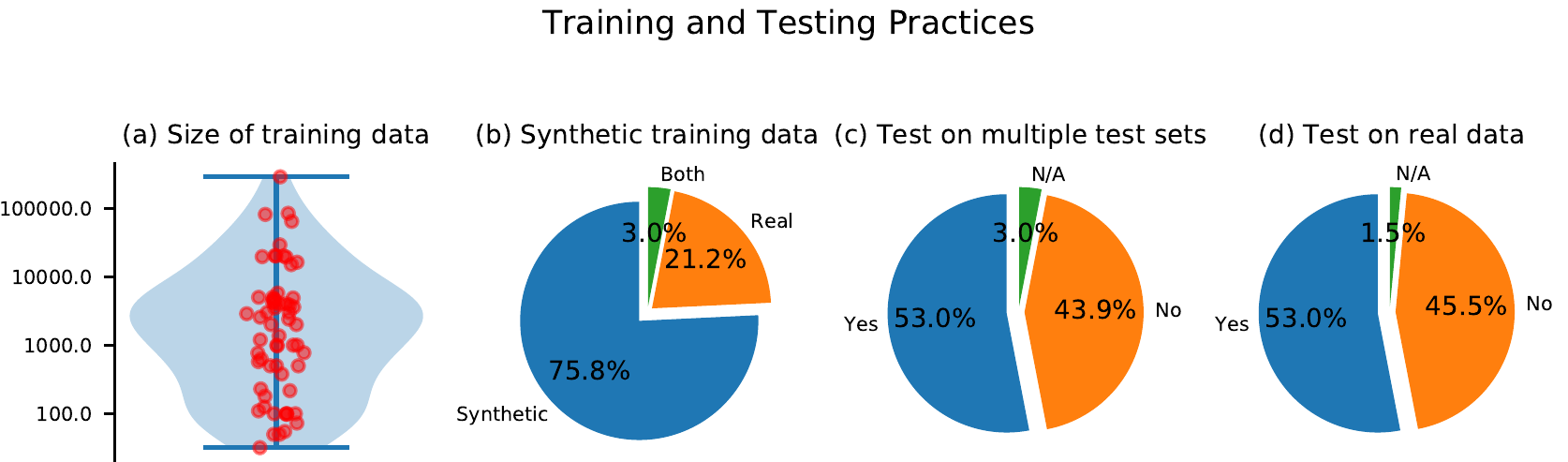}
    \caption{Analysis of the data used in the reviewed papers. (a) shows the distribution of the number of samples in the training data set, (b) shows the percentage of papers working with synthetic or experimental training data, (c) shows the percentage of papers that tested their approaches on multiple data sets including test data from a data distribution different than the training data and (d) shows how many papers tested their approach on real data.}
    \label{fig:data}
\end{figure}

\textbf{Data.} Data is key to successfully apply machine learning techniques to any given problem. We analyzed the usage of data in the reviewed papers and summarized the findings in Fig \ref{fig:data}.\\

\emph{Training.} The number of training data ranged from 32 to 296.300 samples with a median number of training samples of 2.400. As evident from these findings, one of the core bottlenecks of the application of deep learning algorithms to \ac{PAI} is the lack of reliable experimental training data. This can in particular be caused by a lack of ground truth information on the underlying optical tissue properties or the underlying initial pressure distribution when acquiring experimental measurements. To address this issue, researchers make heavy use of simulated data and as a matter of fact, nearly 75\% of papers relied exclusively on these for training the neural network. Table \ref{tab:train_test} shows the distribution of papers that use experimental data. The table shows that the lack of experimental training data is particularly emphasized for the optical and acoustic inverse problem. In contrast to the other tasks, where manual image annotations can be used as a ground truth reference, the underlying initial pressure distribution or optical tissue properties are generally not known in experimental settings. We have identified three main strategies for generating synthetic training data in this review: random, model-based, and reference-based data generation:

\begin{enumerate}
    \item \emph{Random data generation.} The first and simplest strategy generates data by creating completely random distributions of the optical and acoustic properties that are necessary for the simulation framework\cite{cai_end--end_2018}. Here, usually, a Gaussian distribution of the parameters in question is assumed and no dedicated structures are added to the data.
    \item \emph{Model-based data generation.} Training data is created by defining geometrical structures that are assigned optical and acoustic properties according to a hand-crafted model\cite{grohl_estimation_2019}. Such a model might include literature references e.g. for the size, shape, and properties of typical absorbers in tissue. For the generation of training data, many different instances of the model are created that all yield different distributions of chromophores.
    \item \emph{Reference-based data generation.} For the reference-based approach, reference images of different imaging modalities are taken as the basis for data generation\cite{hauptmann_model-based_2018}. They are processed in a way that allows for their direct usage to either create distinct segmentation patterns of, for example, vessels or as the initial pressure distribution for subsequent acoustic forward modeling.
\end{enumerate}

Naturally, researchers also utilized combinations of these approaches, including training on a large data set of simulated data and utilizing a smaller experimental data set to adjust the neural network to the experimental data distribution in a process called \emph{transfer learning}\cite{hauptmann_model-based_2018, pan2009survey}.\\

\begin{table}[h!tb]
    \centering
    \begin{tabular}{lccc}
         \textbf{Problem} & \textbf{Exp. test data} & \textbf{Exp. train data} \\
         \hline
         \\
         Acoustic inverse problem & 8 (40\%) & 1 (5\%)\\
         Image post-processing & 14 (70\%) & 7 (35\%)\\
         Optical inverse problem & 1 (33\%) & 1 (33\%)\\
         Semantic image annotation & 12 (52\%) & 9 (39\%)\\
         \\
    \end{tabular}
    \caption{Overview of the findings for training and test data used in the reviewed papers. The table shows the absolute and relative number of papers that use experimental data for testing or for training.}
    \label{tab:train_test}
\end{table}

\emph{Testing.} In the field of medical imaging, only few prospective studies warrant reliable insights into the fidelity of deep learning methods \cite{liu2019comparison}. One of the major problems is that algorithms are not directly usable by clinicians due to technical or bureaucratic limitations \cite{panch2019inconvenient}. Given the fact that most approaches use simulated data to train their algorithms, there is a high probability that many of the presented algorithms - while yielding superb results on the publication data - could fail in a clinical scenario. This can be attributed to the fact that training data can suffer from several shortcomings compared to the data distribution in reality, such as a significant difference in the data (domain gap)\cite{ross2018exploiting}, an insufficient number of samples (sparsity)\cite{cciccek20163d}, or a selection bias\cite{kato2018learning}. Fig \ref{fig:data} shows that in \ac{PAI} 50\% of papers tested their deep learning approaches on multiple data sets that are significantly different from the training data distribution. Nearly all of these papers test their approaches on experimental data, and about 25\% of the examined papers test on \emph{in vivo} data.\\

\textbf{Simulation frameworks.} Given the necessity to create synthetic data sets for algorithm training, it is crucial to realistically simulate the physical processes behind \ac{PAI}. To this end, we have identified several eminent open-source or freely available frameworks that are being utilized in the field and briefly present five of them here:\\

1) The \emph{k-Wave}\cite{treeby2010k} toolbox is a third-party MATLAB toolbox for the simulation and reconstruction of PA wave fields. It is designed to facilitate realistic modeling of \ac{PAI} including the modeling of detection devices. As of today it is one of the most frequently used frameworks in the field and is based on a k-space pseudo-spectral time-domain solution to the PA equations.\\

2) The \emph{mcxyz} \cite{jacques2014coupling} simulation tool uses a Monte Carlo model of light transport to simulate the propagation of photons in heterogeneous tissue. With this method, the absorption and scattering properties of tissue are used to find probable paths of photons through the medium. The tool uses a \ac{3D} Cartesian grid of voxels and assigns a tissue type to each voxel, allowing to simulate arbitrary volumes.\\

3) The \emph{Monte Carlo eXtreme} \cite{fang2009monte} (MCX) tool is a Graphics Processing Units (GPU)-accelerated photon transport simulator. It is also based on the Monte Carlo model of light transport and supports the simulation of arbitrarily complex \ac{3D} volumes using a voxel domain, but is also capable of simulating photon transport for \ac{3D} mesh models (in the MMC version). Its main advantage is the support of GPU acceleration using a single or multiple GPUs.\\

4) The \emph{NIRFAST} \cite{dehghani2009near} modeling and reconstruction package was developed to model near-infrared light propagation through tissue. The framework is capable of single-wavelength and multi-wavelength optical or functional imaging from simulated and measured data. It recently integrated the NIRFAST optical computation engine into a customized version of 3DSlicer.\\

5) The \emph{Toast++}\cite{schweiger2014toast++} software suite consists of a set of libraries to simulate light propagation in highly scattering media with heterogeneous internal parameter distribution. Among others, it contains numerical solvers based on the finite-element method, the discontinuous Galerkin discretization scheme, as well as the boundary element method. \\

\textbf{Neural network Architectures.} \acp{CNN}\cite{gu2018recent} are currently the state-of-the-art method in deep learning-based \ac{PAI}. Here, especially the U-Net\cite{ronneberger2015u} architecture is highly popular and has been used in over 70\% of the reviewed papers. Other architectures such as residual neural networks or fully-connected networks only exhibit a prevalence of less than 10\% each. It should be noted that usually, slightly modified versions of the vanilla U-Net, especially fitting to the target application, have proven to yield the best performance.\\

\section{Acoustic Inverse Problem}
\label{sec:reconstruction}  

The acoustic inverse problem refers to the task of reconstructing an image of the initial pressure distribution from measured time-series pressure data. Reconstructing a PA image from time-series data constitutes the main body of work, either by \emph{enhancing existing model-based approaches} (40\% of papers) or by performing \emph{direct image reconstruction} (35\% of papers). Furthermore, auxiliary tasks are being examined as well, such as the \emph{localization of wavefronts} from point sources (15\% of papers) and the \emph{estimation of the speed of sound} of the medium (10\% of papers). Information on these parameters is important to achieve optimal image reconstruction, thereby enhancing the image quality and improving the image's usefulness in clinical scenarios. Typical evaluation metrics that are used to assess the performance of reconstruction algorithms are the \ac{MSE}, \ac{MRE}, \ac{MAE}, \ac{PSNR}, \ac{SSIM} Coefficient, \ac{CC}, and \ac{SNR}. In total, we identified 20 papers that tackle the acoustic inverse problem, all of which use simulated PA data for training. Surprisingly, approximately 40\% of papers presented results on experimental data either using phantoms or \emph{in vivo} (animal or human) measurements. In the following, we summarize the literature partitioned into the already mentioned sub-topics: deep learning-enhanced model-based image reconstruction, direct image reconstruction, point source localization, and speed of sound estimation.\\

\subsection{Deep learning-enhanced model-based image reconstruction}

The central idea is to leverage the flexibility of deep learning to enhance already existing model-based reconstruction algorithms\cite{schwab_deep_2019,schwab_real-time_2018}, by introducing learnable components. To this end, Schwab \emph{et al.}\cite{schwab_learned_2019} proposed an extension of the weighted universal back-projection algorithm. The core idea is to add additional weights to the original algorithm, with the task of the learning algorithm then being to find optimal weights for the reconstruction formula. By learning the weights, the authors were able to reduce the error introduced from limited view and sparse sampling configurations by a factor of two. Furthermore, Antholzer \emph{et al.}\cite{antholzer_nett_2019} and Li \emph{et al.}\cite{li_nett_2020} leveraged neural networks to learn additional regularization terms for an iterative reconstruction scheme. Hauptmann \emph{et al.}\cite{hauptmann_model-based_2018} demonstrated the capability for \acp{CNN} to perform iterative reconstruction by training a separate network for each iteration step and integrating it into the reconstruction scheme. The authors showed that since their algorithm was trained on synthetic data, several data augmentation steps or the application of transfer learning techniques were necessary to achieve satisfactory results. Finally, Yang \emph{et al.}\cite{yang_accelerated_2019} demonstrated the possibility to share the network weights across iterations by using recurrent inference machines for image reconstruction.\\

\emph{Key Insights:} An interesting insight shared in one of the papers by Antholzer \emph{et al.}\cite{antholzer_nett_2019} was that model-based approaches seem to work better for "exact data", while deep learning-enhanced methods outperform purely model-based approaches on noisy data. This makes the application of deep learning techniques very promising for the typically noisier and artifact-fraught experimental data\cite{kazakeviciute2016multispectral,manwar2018photoacoustic}. On the other hand, currently employed deep learning models do not seem to generalize well from simulated to experimental data as evident from the fact that only 40\% of papers tested their method on experimental data (cf. table \ref{tab:train_test}). Ideally, the algorithms would have to be trained on experimental data.\\

\subsection{Direct image reconstruction}

\begin{figure}[h!tb]
    \centering
    \includegraphics{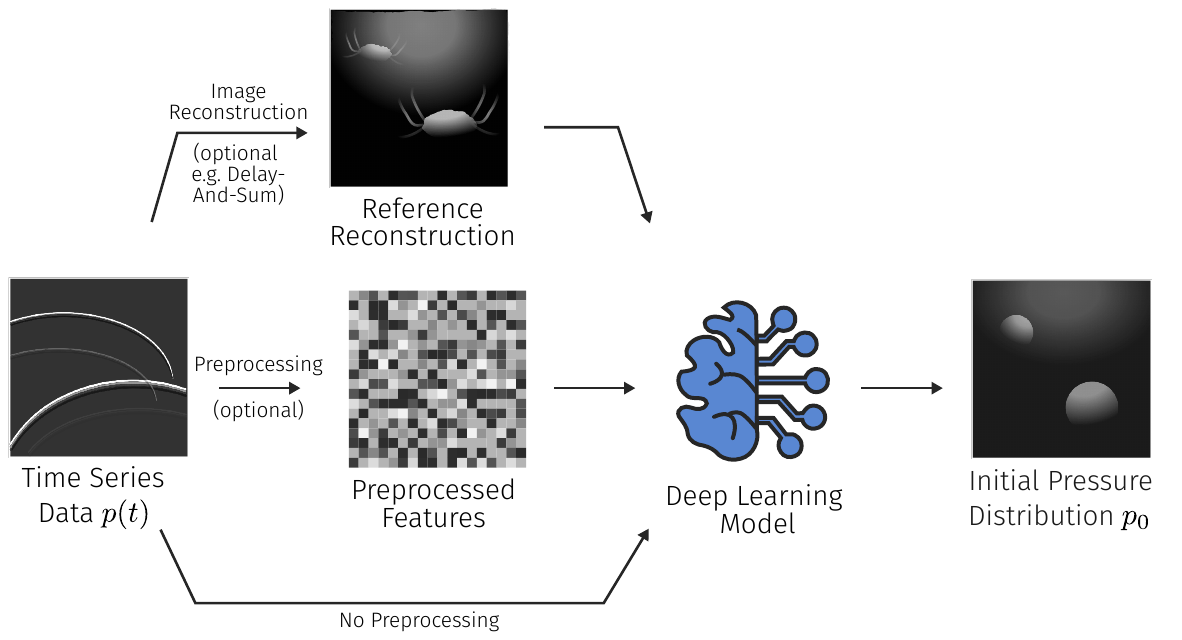}
    \caption{Visualization of the principal approaches to deep learning-based PA image reconstruction. The time-series data is either given directly to a neural network, or after preprocessing steps, such as reference reconstructions or the calculation of hand-crafted feature maps. The goal of the reconstruction is to estimate the underlying initial pressure distribution.}
    \label{fig:aip:imageReconstruction}
\end{figure}

The principal idea of direct image reconstruction with deep learning is to either completely replace the commonly used model-based methods or to integrate model-based methods as additional information for a deep learning-based solution. An overview  of these approaches are summarized in Fig \ref{fig:aip:imageReconstruction}.\\

The first approaches to direct image reconstruction with \acp{CNN} were proposed in 2018 by Waibel \emph{et al.}\cite{waibel_reconstruction_2018} and Anas \emph{et al.}\cite{anas_robust_2018}. Waibel \emph{et al.}\cite{waibel_reconstruction_2018} and Lan \emph{et al.}\cite{lan_reconstruct_2019} used modified U-Net architectures to estimate the initial pressure distribution directly from time-series pressure data, whereas Anas \emph{et al.}\cite{anas_robust_2018} used a \ac{CNN} architecture with dense blocks. Furthermore, Lan \emph{et al.} \cite{lan_ki-gan_2019,lan_hybrid_2019} proposed a method based on a generative adversarial network\cite{goodfellow2014generative} approach that - in addition to time-series data - also uses a reconstructed PA image as additional information to regularize the neural network. Guan \emph{et al.} \cite{guan_limited_2019} compared implementations of all these techniques to assess their merit in brain imaging within a neurological setting. They compared an algorithm that directly estimates the reconstructed image from time-series data, a post-processing approach, as well as a custom approach with hand-rafted feature vectors for the model. Their results show that adding additional information improves the quality of the reconstructed image, that iterative reconstruction generally worked best for their data, and that deep learning-based reconstruction was faster by 3 orders of magnitude.\\

\emph{Key Insights:} In contrast to deep learning-enhanced model-based reconstruction, direct deep learning reconstruction schemes are comparatively easy to train and most of the papers utilize the U-Net as their base architecture. In several works it was demonstrated that the \emph{infusion of knowledge} by regularizing the network with reference reconstructions or additional data from hand-crafted preprocessing steps led to very promising results\cite{lan_ki-gan_2019,lan_hybrid_2019}, generalizing them in a way that led to first successes on \emph{in vivo} data. Considering that deep learning-based image reconstruction outperforms iterative reconstruction techniques in terms of speed by orders of magnitude\cite{guan_fully_2019}, it is safe to say that these methods can be a promising avenue of further research. It has to be noted that, especially in terms of robustness and uncertainty-awareness, the field has much room for improvement. For example, Sahlstr\"om \emph{et al.} \cite{sahlstrom2020modeling} have modeled uncertainties of the assumed positions of the detection elements for model-based image reconstruction, but no comparable methods were applied in deep learning-based \ac{PAI} as of yet.\\

\subsection{Point source localization}

\begin{figure}[h!tb]
    \centering
    \includegraphics{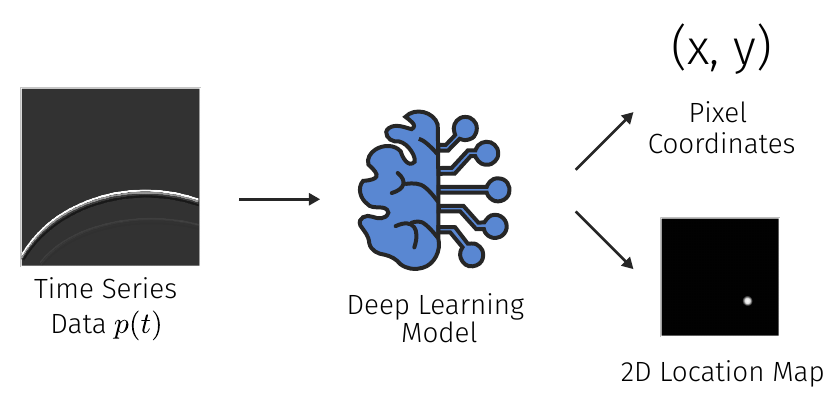}
    \caption{Approaches for point source localization use time-series data as input to estimate either the pixel coordinates of the point of origin of the pressure wave or a heat map containing the probability of the source being in a certain location of the image.}
    \label{fig:aip:pointSourceLocalization}
\end{figure}

The localization of the spatial position of point sources from time-series PA measurements was identified as a popular sub-task concerning PA image reconstruction. An algorithm for this could for example be used for the automatic detection and localization of point absorbers, such as needle tips, in a PA image. The general idea is to take time-series data to either regress numerical values for the pixel coordinates of the sources of the wavefronts or to output a two-dimensional map of the possible source locations (see Fig. \ref{fig:aip:pointSourceLocalization}).\\

To this end, Reiter \emph{et al.}\cite{reiter_machine_2017} presented an approach that uses a \ac{CNN} to transform the time-series data into an image that identifies the \ac{2D} point localization of the wavefront origin. They further use this approach to distinguish between signals and artifacts in time-series data. Johnstonbaugh \emph{et al.} \cite{johnstonbaugh_novel_2019} also use a \ac{CNN} in an encoder-decoder configuration to reconstruct the PA signal into an image containing a single point source. A similar architecture proposed by the same group \cite{johnstonbaugh_deep_2020} is also used to process the time-series data and output cartesian coordinates of the point source location.\\

\emph{Key Insights:} Similar to the deep learning-based direct reconstruction methods, methods for point source localization are exceptionally easy to train and can even be trained on \emph{in vitro} experimental data. This ease of accessibility made this task the first application of deep learning in \ac{PAI} \cite{reiter_machine_2017}. However, the integrability of these methods into clinical practice and their future impact beyond certain niche applications is questionable because \emph{in vivo} scenarios do typically not exclusively consist of point sources but comprise a very complex and heterogeneous distribution of chromophores.

\subsection{Speed of sound estimation}

A correct estimate of the speed of sound within the medium is an important constituent to successful image reconstruction. We identified two papers that explicitly incorporated the estimation of the speed of sound into their reconstruction. Shan \emph{et al.} \cite{shan_simultaneous_2019} used a \ac{CNN} to reconstruct the initial pressure distribution as well as the speed of sound simultaneously from the time-series data and Jeon \emph{et al.}\cite{jeon_deep_2020} trained a U-Net to account for the speed of sound aberrations that they artificially introduced to their time-series data.\\

\emph{Key Insights:} Automatically integrating estimates of the speed of sound into the image reconstruction algorithm can substantially enhance image quality and hence is an interesting direction of research. Nevertheless, the formulation of a corresponding optimization problem is inherently difficult, as it is not straightforward to assess the influence of a speed of sound mismatch on a reconstruction algorithm. Furthermore, the validation of these methods is difficult, as there typically is no \emph{in vivo} ground truth available.

\section{Image Post-Processing}
\label{sec:processing}

\begin{figure}[h!tb]
    \centering
    \includegraphics{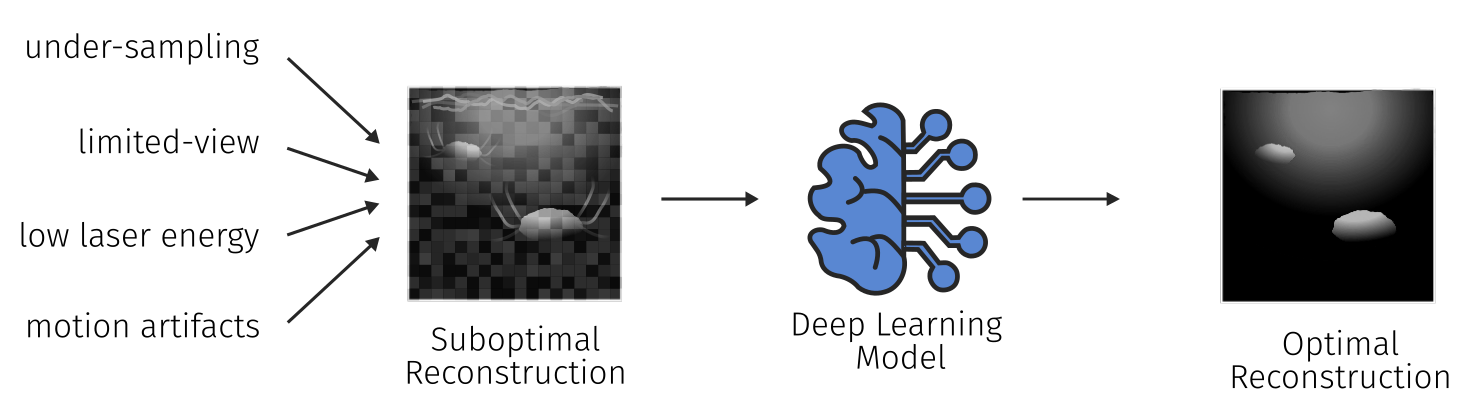}
    \caption{Post-processing techniques are tasked to improve the image quality of a reconstructed PA image. The image quality can be reduced by many factors including under-sampling, limited-view artifacts, low laser energy, or the presence of motion during the measurement.}
    \label{fig:postprocessing}
\end{figure}

Being a comparatively young imaging modality, \ac{PAI} still suffers from distinct artifacts\cite{kazakeviciute2016multispectral}. These can have multiple origins and are primarily caused by hardware limitations such as light absorption in the transducer membrane or fluctuations in the pulse laser energy\cite{manwar2018photoacoustic}. Other issues can also lead to decreased image quality, such as under-sampling or limited-view artifacts, as well as other influences such as motion artifacts or artifacts specific to the reconstruction algorithm (see Fig \ref{fig:postprocessing}). Research in the field of using post-processing algorithms can broadly be divided into two main areas: the elimination of artifacts  \ref{sec:processing:artifact} which mostly encompass systematic error sources and the enhancement of image quality \ref{sec:processing:quality} which is lost mainly through stochastic error sources.\\

\subsection{Artifact removal}
\label{sec:processing:artifact}

One principal approach to speed up image reconstruction is to use sparse data that only contains a fraction of the available time-series data. While this potentially leads to a significant increase in reconstruction speed, it comes with a cost in form of the deterioration of the image quality and the introduction of characteristic under-sampling artifacts. Several groups\cite{davoudi_deep_2019,antholzer_deep_2019,guan_fully_2019,farnia2020high} have shown that a large portion of these artifacts can be recovered using deep learning techniques. A core strength of such approaches is that experimental PA data can be utilized for training, by artificially undersampling the available channels and training the algorithm to predict the reconstructions from (1) full data, (2) sparse data, or (3) limited-view data \cite{antholzer_photoacoustic_2018,antholzer_deep_2018,deng_machine-learning_2019,zhang2020new,godefroy2020solving,tong2020domain}.\\

Reflection artifact can be introduced by the presence of acoustic reflectors in the medium (for example air). Allman \emph{et al.}\cite{allman_photoacoustic_2018} showed that deep learning can be used to distinguish between artifacts and true signals and Shan \emph{et al.}\cite{shan_accelerated_2019} demonstrated that the technology is also capable of removing such artifacts from the images. Furthermore, Chen \emph{et al.}\cite{chen_deep-learning-based_2019} introduced a deep learning-based motion correction approach for PA microscopy images that learns to eliminate motion-induced artifacts in an image.\\

\emph{Key Insights:} For limited-view or limited-data settings, experimental training data can easily be created by artificially constraining the available data, for example, by under-sampling the number of available time series data. On the other hand, for the task of artifact removal, it can be comparatively difficult to train models on \emph{in vivo} experimental settings for different sources of artifacts. This is, because artifacts can have various origins and are also dependent on the specific processing steps. Nevertheless, impressive results of the capability of learning algorithms to remove specific artifacts were demonstrated.

\subsection{Image quality enhancement}
\label{sec:processing:quality}

The quality and resolution of PA images are also limited by several other factors including the limited bandwidth of PA detectors, the influence of optical and acoustic scattering, the presence of noise due to the detection hardware, and fluctuations in the laser pulse-energy.\\

To remedy this, Gutta \emph{et al.}\cite{gutta_deep_2017} and Awasthi \emph{et al.}\cite{awasthi_deep_2020} proposed methods that aim to recover the full bandwidth of a measured signal. This is achieved by obtaining pairs of full bandwidth and limited bandwidth data using simulations that are used to train a neural network. Since experimental systems are always band-limited, the authors of these works rely on the presence of simulated data. On the other hand, more classical deep learning-based super-resolution algorithms were proposed by Zhao \emph{et al.}\cite{zhao_new_2020} to enhance the resolution of PA devices in the context of PA microscopy. For training of super-resolution approaches, the authors are theoretically not restricted by the domain of application and as such can also use data from sources unrelated to \ac{PAI}.\\

Several approaches have been proposed to enhance the image quality by improving the signal-to-noise ratio of image frames acquired with low energy illumination elements, such as LED-based systems. This has generally been done using \acp{CNN} to improve a single reconstructed image, for example by Vu \emph{et al.}\cite{vu_generative_2020}, Singh \emph{et al.}\cite{singh_deep_2020}, Anas \emph{et al.}\cite{anas_enabling_2018}, and Hariri \emph{et al.} \cite{hariri_deep_2020} or by using a neural network to fuse several different reconstructions into a higher-quality version, as proposed by Awasthi \emph{et al.}\cite{awasthi_pa-fuse_2019}.\\

\emph{Key Insights:} For the enhancement of image quality, common deep learning tasks from the field of computer vision\cite{voulodimos2018deep} can be translated to PA images relatively easily, as the algorithms are usually astonishingly straightforward to train and validate. We believe that applying well-established methods from fields adjacent to \ac{PAI} can be of excellent benefit to the entire field.

\section{Optical Inverse Problem}
\label{sec:quantification}

The optical inverse problem is concerned with estimating the optical tissue properties from the initial pressure distribution. The first method proposed to solve this inverse problem was an iterative reconstruction scheme to estimate the optical absorption coefficient\cite{cox2006two}. Over time, the iterative inversion schemes have become more involved\cite{cox2010quantitative} and Buchmann \emph{et al.}\cite{buchmann2020quantitative} achieved first successes towards experimental validation. Recently, data-driven approaches for the optical inverse problem have emerged, including classical machine learning\cite{kirchner2018context} as well as deep learning approaches. A tabulated summary of the identified papers can be found in Table \ref{tab:summary:oip} \\

\begin{table}[h!tb]
    \centering
    \begin{tabular}{lp{3cm}p{2cm}p{2cm}p{2cm}p{2cm}}
         \textbf{Publication} & \textbf{Base \mbox{Architecture}}  & \textbf{target $\mu_a$ $[cm^{-1}]$} & \textbf{background $\mu_a$ $[cm^{-1}]$} & \textbf{background $\mu_s'$ $[cm^{-1}]$} & \textbf{Validation Data}\\
         \hline
         \\
         Cai \emph{et al.} \cite{cai_end--end_2018} & U-Net with residual blocks  & N/A & $0.2 - 0.4$ & $5 - 10$ & \emph{in silico} \\
         Gr{\"o}hl \emph{et al.} \cite{grohl_confidence_2018} & U-Net & $2 - 10$ & const. $0.1$ & const. $1.5 $ & \emph{in silico} \\
         Chen \emph{et al.} \cite{chen_deep_2020} & U-Net & $0.1 - 2$ & $0.1 - 0.4$ & const. $10 $ & \emph{in vitro}\\
         \\
    \end{tabular}
    \caption{Tabulated overview of the identified literature regarding the optical inverse problem. The table shows the publication, the base network architecture, the range of absorption and scattering parameters used for the training data, and the type of data that the approach was validated with.}
    \label{tab:summary:oip}
\end{table}

\begin{figure}[h!tb]
    \centering
    \includegraphics{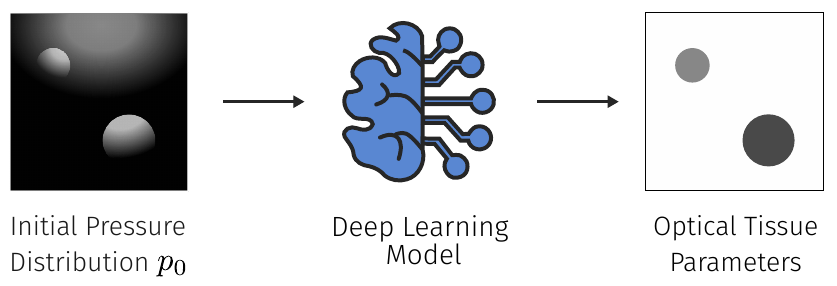}
    \caption{To solve the optical inverse problem, a neural network is tasked to estimate the underlying optical tissue parameters, primarily the optical absorption coefficient, from the initial pressure distribution $p_0$.}
    \label{fig:oip:overview}
\end{figure}

For the identified deep learning-based approaches, the key objective is to estimate the optical absorption coefficients and subsequently the absolute concentrations of chromophores from the initial pressure distribution (see Fig. \ref{fig:oip:overview}).\\

Cai \emph{et al.} \cite{cai_end--end_2018} trained a U-Net with residual blocks to estimate the absolute concentration of \ac{ICG} alongside the relative ratio of $Hb$ and $HbO_2$. To this end, they simulated random smoothed maps of optical tissue properties for training and tested their approach on several simulated data sets, including one created from a digital mouse model\cite{dogdas2007digimouse}. Gr{\"o}hl \emph{et al.} \cite{grohl_confidence_2018} trained a total of four U-Net models to estimate fluence and optical absorption from the initial pressure distribution as well as directly from time-series pressure data. They also presented a method to estimate the expected error of the inversion, yielding an indicator for the model uncertainty. Their approach was trained and tested on simulated data, which contained tubular structures in a homogeneous background. Finally, Chen \emph{et al.}\cite{chen_deep_2020} trained a U-Net to directly estimate optical absorption from simulated images of initial pressure distribution. They trained and tested their model on synthetic data comprising geometric shapes in a homogeneous background, as well as another model on experimental data based on circular phantom measurements.\\

\emph{Key Insights:} Model-based methods to tackle the optical inverse problem suffer from the fact that many explicit assumptions have to be made that typically do not hold in complex scenarios\cite{kirchner2018context}. With data-driven approaches, many of these assumptions are only implicitly made within the data distribution, leaving room for a substantial improvement. Obtaining ground truth information on the underlying optical tissue properties \emph{in vivo} can be considered impossible and is exceptionally involved and error-prone even \emph{in vitro}\cite{grohl_confidence_2018}. As such, there has been no application yet to \emph{in vivo} data, leaving the optical inverse problem as one of the most challenging problems in the field of \ac{PAI}, which is reflected by the comparatively low amount of published research on this topic.\\

\section{Semantic Image Annotation}
\label{sec:clinical_translation}

\begin{figure}[h!tb]
    \centering
    \includegraphics{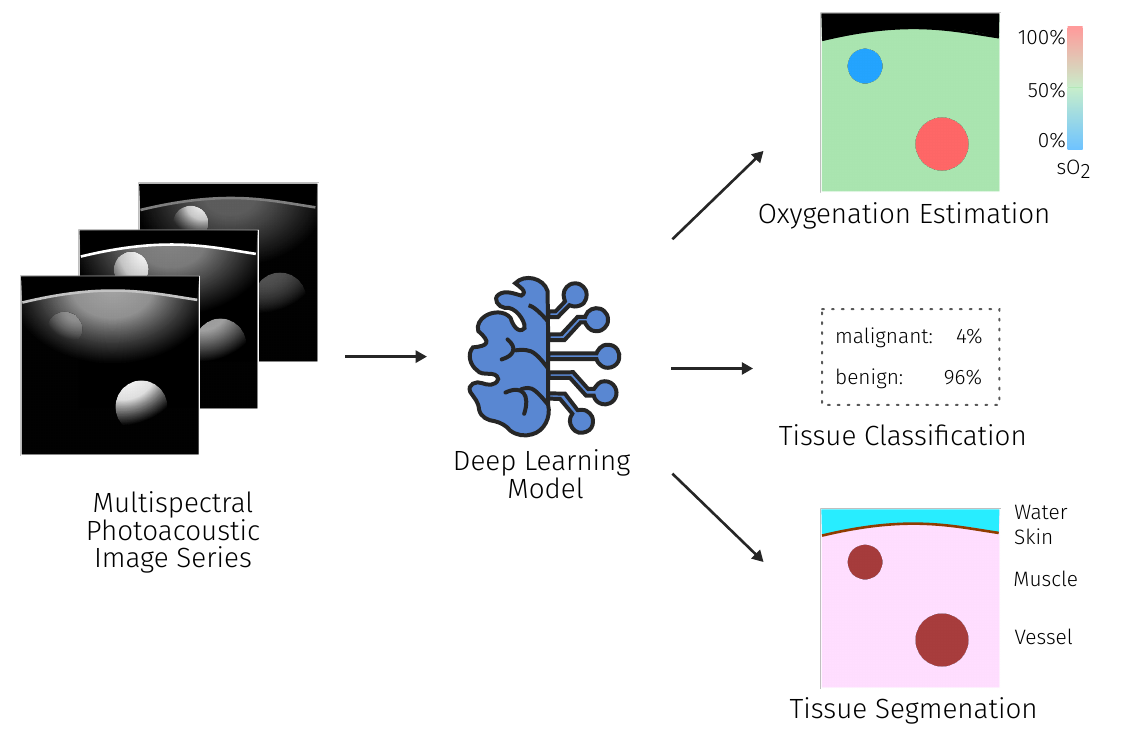}
    \caption{For semantic tissue annotation (typically multispectral) PA measurements are used as the input and the algorithm is tasked to estimate the desired parameters, such as tissue oxygenation or segmentation maps of different tissue types. The black color in the oxygenation estimation denotes areas where oxygenation values cannot be computed.}
    \label{fig:clinical_translation}
\end{figure}

While the topical areas until now have mostly considered PA data at a single wavelength, the power of \ac{PAI} for clinical use cases lies in its ability to discern various tissue properties through analysis of the changes in signal intensity over multiple wavelengths (see Fig \ref{fig:clinical_translation}). This allows for the estimation of functional tissue properties, especially blood oxygenation (section \ref{sec:clinical_translation:functional}), but also for the classification and segmentation (section \ref{sec:clinical_translation:classification}) of tissues and tissue types.\\

\subsection{Functional property estimation}
\label{sec:clinical_translation:functional}

The estimation of local blood oxygenation  $sO_2$ is one of the most promising applications of \ac{PAI}. In principle, information on the signal intensities of at least two wavelengths are needed to unmix the relative signal contributions of oxyhemoglobin $HbO_2$ and deoxyhemoglobin $Hb$: $sO_2 = HbO_2 / (HbO_2 + Hb)$. For \ac{PAI}, wavelengths around the isosbestic point ($\approx 800\,\text{nm}$) are commonly chosen for this task. Because linear unmixing should not directly be applied to the measured signal intensities due to the non-linear influence of the fluence, a lot of work has been conducted to investigate the applicability of neural networks to tackle this problem. Due to the unavailability of ground truth oxygenation information, the networks are currently being trained exclusively on simulated data. The problem was approached using fully-connected neural networks\cite{grohl_estimation_2019} as well as \acp{CNN}\cite{bench2020toward}.\\

The use of feed-forward fully-connected neural networks was demonstrated by Gr\"ohl \emph{et al.}\cite{grohl_estimation_2019} to be capable to yield accurate oxygenation estimations \emph{in silico} from single-pixel $p_0$ spectra. In addition, it was demonstrated that the application of the method to experimental \emph{in vitro} and \emph{in vivo} data yielded plausible results as well. Olefir \emph{et al.}\cite{olefir2020deep} demonstrated that introducing prior knowledge to the oxygenation estimation can improve performance. Specifically, the authors introduced two sources of information for regularization. On the one hand, they introduced fluence eigenspectra which they obtained from simulated data and on the other hand, they also estimated their results based on spectra from a larger patch of tissue thus introducing spatial regularization. They demonstrated the applicability of the method to \emph{in vitro} and \emph{in vivo} data in several experiments.\\

To make full use of the spatial context information, \acp{CNN} were employed to estimate blood oxygenation using the spectra of entire \ac{2D} images rather than single-pixel spectra. This was demonstrated by Yang \emph{et al.}\cite{yang_eda-net_2019,yang_quantitative_2019}, Luke \emph{et al.}\cite{luke_o-net_2019}, and Hoffer-Hawlik \emph{et al.}\cite{hoffer-hawlik_abso2luteu-net_2019}. Furthermore, Bench \emph{et al.}\cite{bench2020toward} showed the feasibility to estimate oxygenation from multispectral \ac{3D} images. It has to be noted that there exists a trade-off regarding the spatial extent and number of voxels of the input images and the number of training images that can feasibly be simulated for algorithm training. The approaches demonstrate the feasibility of using \acp{CNN} for estimating oxygenation with high accuracy (for reported values in the publication see Table \ref{tab:results_oxygenation_estimation}), however, a successful application of these methods \emph{in vitro} or \emph{in vivo} has not yet been shown, which is most probably caused by the large domain gap between simulated and experimental PA images.\\

\begin{table}[h!tb]
    \centering
    \begin{tabular}{ll}
        \textbf{Publication} & \textbf{Reported sO$_2$ estimation error} \\
        \hline
        \\
        Bench \emph{et al.}\cite{bench2020toward} & 4.4\% $\pm$ 4.5\% MAE\\
        Cai \emph{et al.} \cite{cai_end--end_2018} & 0.8\% $\pm$ 0.2\% MRE\\
        Gr\"ohl \emph{et al.}\cite{grohl_estimation_2019} & 6.1\% MedRE, IQR: [2.4\%, 18.7\%]\\
        Hoffer-Hawlik \emph{et al.}\cite{hoffer-hawlik_abso2luteu-net_2019} & RSME consistently below 6\%\\
        Luke \emph{et al.}\cite{luke_o-net_2019} & 5.1\% MedAE at 25dB SNR\\
        Olefir \emph{et al.}\cite{olefir2020deep} & 0.9\%, IQR [0.3\%, 1.9\%] to 2.5\%, IQR [0.5\%, 3.5\%] MedAE*\\ 
        Yang \emph{et al.}\cite{yang_quantitative_2019} & 1.4\% $\pm$ 0.2\% MRE \\
        Yang and Gao \cite{yang_eda-net_2019} & 4.8\% $\pm$ 0.5\% MAE\\
        \\
    \end{tabular}
    \caption{Overview of some of the reported errors on sO$_2$ estimation. Standard deviations or interquartile ranges (IQR) are shown if reported. It has to be noted that the used metrics as well as the conventions which pixels the error metrics are calculated on vary drastically between papers. As such the numbers are not directly comparable. For more detailed results, please refer to the linked papers. MedRE = Median Relative Error; MedAE = Median Absolute Error; MRE = Mean Relative Error; MAE = Mean Absolute Error; RSME = Root Mean Squared Error; *depending on dataset.}
    \label{tab:results_oxygenation_estimation}
\end{table}

The estimation of other functional tissue properties has also been investigated, such as the detection of glucose by Ren \emph{et al.}\cite{ren_effects_2018}, the size of fat adipocytes by Ma \emph{et al.}\cite{ma_adipocyte_2018}, as well as the unmixing of arbitrary chromophores in an unsupervised manner by Durairaj \emph{et al.}\cite{durairaj_unsupervised_2020}.\\

\emph{Key Insights:} The estimation of functional tissue properties is closely related to the optical inverse problem, as functional properties can be derived from the optical properties of the tissue. But the direct estimation of the desired properties without quantification of the optical properties in an intermediate step is very popular. One reason for this is that there exist reference methods that can measure the functional properties and can be used to validate the results\cite{olefir2020deep}. This potentially also enables training an algorithm on experimental data, when using reference measurements as the ground truth. Taking the estimation of tissue oxygenation as an example showcases the potential rewards of comprehensively solving this family of problems, as it would enable a lot of promising applications, such as oxygen-level specific dosimetry in radiotherapy\cite{matsuo2014magnetic} or cancer type classification based on local variations in blood oxygenation in the tumor's microenvironment\cite{horsman1998measurement}.

\subsection{Tissue classification and segmentation}
\label{sec:clinical_translation:classification}

Multispectral image information can also be used to differentiate between different tissue types or to identify and detect tissue pathologies. In such cases, strategies for dataset curation differ depending on the use case but using experimental datasets is possible with manual data annotation. In the work of Moustakidis \emph{et al.} \cite{moustakidis_fully_2019} \emph{in vivo} images of a raster-scan optoacoustic mesoscopy (RSOM) system were utilized to automatically differentiate between different skin structures, while Lafci \emph{et al.} \cite{lafci_efficient_2020} used neural networks to segment the entire imaged object. Furthermore, Wu \emph{et al.}\cite{wu_multi-wavelength_2017} imaged \emph{ex vivo} tissue samples to monitor lesion formation during high-intensity focused ultrasound (HIFU) therapy, and Jnawali \emph{et al.}\cite{jnawali_deep_2019, jnawali_transfer_2019, jnawali_automatic_2020} also analyzed \emph{ex vivo} tissue to differentiate cancer tissue from normal tissue in pathology samples.\\

On the other hand, we identified several papers that used simulated data to train their networks. Typically, simulations of acoustic waves are conducted using pre-processed images of a different modality, such as CT images, and treating the intensity distribution as the initial pressure distribution. This was done by Zhou \emph{et al.}\cite{zhou_analysis_2019} to investigate the feasibility to differentiate healthy bone tissue from pathologies such as hyperosteogeny and osteoporosis. Further work by Zhang \emph{et al.}\cite{zhang_photoacoustic_2018} also examined the feasibility of DL-based breast cancer classification, Lin \emph{et al.}\cite{lin_computer-aided_2019} investigated the feasibility of endometrial cancer detection, and several groups including Zhang \emph{et al.}\cite{zhang_pathology_2018}, Luke \emph{et al.}\cite{luke_o-net_2019}, Chlis \emph{et al.}\cite{chlis_sparse_2019}, and Boink \emph{et al.}\cite{boink_partially-learned_2019} examined the segmentation of blood vessels. Finally, Allman et al.\cite{allman_deep_2019} conducted feasibility experiments that demonstrated the capability of neural networks to automatically segment needle tips in PA images.\\

\emph{Key Insights:} Semantic image annotation enables intuitive and fast interpretation of PA images. Given the number of potential applications of \ac{PAI}, we believe that semantic image annotation is a promising future research direction. Because the modality is comparatively young, high-quality reference data for algorithm training that are annotated by healthcare professionals are very rare. Furthermore, the cross-modality and inter-institutional performance of \ac{PAI} devices has to our knowledge not been examined as of yet. This makes validation of the proposed algorithms \emph{in vivo} difficult, as reflected by some of the presented work. As discussed throughout this review, the image quality of \ac{PA} images relies heavily on the solutions to the acoustic and optical inverse problems. This potentially introduces difficulties for manual data annotation and thus makes it more difficult to integrate developed methods into clinical practice.

\section{Discussion}
\label{sec:discussion}

The clinical translation of deep learning methods in \ac{PAI} is still in its infancy. Even though many classical image processing tasks, as well as PA-specific tasks, have already been tackled using deep learning techniques, vital limitations remain. For instance, many researchers resort to simulated data due to the lack of high-quality annotated experimental data. Accordingly, none of the proposed techniques were validated in a large-scale clinical \ac{PAI} study. In this section we will discuss the challenges for clinical translation of deep learning methods in \ac{PAI} and will conclude by summarizing the key findings of this review.\\

The challenges of clinical translation of spectroscopic optical imaging techniques have previously been extensively examined by Wilson \emph{et al.}\cite{wilson2018challenges}. While their work focused primarily on the general challenges and hurdles in translating biomedical imaging modalities into clinical practice, in this review, we focused on the extent to which the application of deep learning in particular could facilitate or complicate the clinical integration of \ac{PAI}. To this end, we have summarized important features that a learning algorithm should fulfill, based on the findings in other literature \cite{wilson2018challenges,miotto2018deep,kelly2019key}:\\

\textbf{Generalizability.} In general, training data \emph{must} be representative of the data encountered in clinical practice to avoid the presence of biases\cite{pannucci2010identifying} in the trained model. The acquisition of high-quality experimental data sets in \ac{PAI} is extremely problematic due to, for example, the high intra-patient signal variability caused by changes in local light fluence, the small amount of clinically approved PAI devices, and the lack of reliable methods to create ground truth annotations.\\

The lack of experimental data can be attributed to the comparative youth of \ac{PAI}, but also to the fact that semantic information for images is only available via elaborate reference measurement setups or manual data annotations, which are usually created by healthcare professionals. But even in commonplace imaging modalities like Computed Tomography (CT) or Magnetic Resonance Imaging (MRI), high-quality annotations are extremely costly, time-consuming and thus sparse compared to the total number of images that are taken. Finally, annotations of optical and acoustic properties are intrinsically not available \emph{in vivo}, as there currently exist no gold standard methods to non-invasively measure, for example, optical absorption, optical scattering, or the speed of sound.\\

To account for the lack of available experimental data, approximately 75\% of models were trained on simulated photoacoustic data. This practice has led to many insights into the general applicability and feasibility of deep learning-based methods. But methods trained purely on simulated data have shown poor performance when being applied to experimental data. Systematic differences between experimental PA images and those generated by computational forward models are very apparent. These differences are commonly referred to as the \textit{domain gap} and can cause methods to fail on \textit{in vivo} data despite thorough validation \textit{in silico}, since deep learning methods cannot easily generalize to data from different distributions. Closing this gap can make the \textit{in silico} validation of deep learning methods more meaningful. We see several approaches to tackle this problem: 

\begin{enumerate}
    \item Methods to create more realistic simulations. This comprises the implementation of digital twins of the respective PA devices or the simulation of anatomically realistic geometric variations of tissue.
    \item Domain adaptation methods that are currently developed in the field of computer vision\cite{voulodimos2018deep} could help translate images from the synthetic to the real PA image domain.
    \item Methods to refine the training process, such as extensive data augmentation specific for \ac{PAI}, the weighting of training data\cite{wirkert2017physiological}, content disentanglement\cite{ilse2020diva} or domain-specific architecture changes\cite{lan_ki-gan_2019}, as well as the tight integration of prior knowledge into the entire algorithm development pipeline\cite{hauptmann_model-based_2018}.
\end{enumerate}

A promising opportunity could lie in the field of \emph{life-long learning}\cite{ruvolo2013ella,chen2018lifelong}. This field strives to develop methods that have the ability to continuously learn over time by including new information while retaining prior knowledge\cite{parisi2019continual}. In this context, research is conducted towards developing algorithms that efficiently adapt to new learning tasks (\emph{meta-learning})\cite{finn2017model} and can be trained on various different but related tasks (\emph{multi-task learning})\cite{caruana1997multitask}. The goal is to create models that can continue to learn from data that becomes available after deployment. We strongly believe that the integration of such methods into the field of \ac{PAI} can have exceptional merit, as this young imaging modality can be expected to undergo frequent changes and innovations in the future.\\

\textbf{Uncertainty estimation.} We strongly believe that methods should be uncertainty-aware, since gaining insight into the confidence of deep learning models can serve to avoid blindly assuming estimates to be accurate\cite{what2017kendall,osawa2019practical}. The primary goal of uncertainty estimation methods is to provide the confidence interval for measurements, for example, by calculating the average and standard deviation over a multitude of estimation samples\cite{gal2016dropout}. On the other hand, such metrics might not be sufficient and a current direction of research is to recover the full posterior probability distribution for the estimate given the input, which for instance enables the automatic detection of multi-modal posteriors\cite{ardizzone2018analyzing}. Uncertainty estimation and Bayesian modeling of the inverse problems is an active field of research in \ac{PAI}\cite{tick2016image,tick2019modelling,sahlstrom2020modeling}. While a first simulation study\cite{grohl_confidence_2018} has demonstrated the benefits of exploiting uncertainty measures when using deep learning methods, the potential of this branch of research remains largely untapped.\\

\textbf{Out-of-distribution detection.} A major risk of employing deep learning-based methods in the context of medical imaging can be seen in their potentially undefined behavior when facing out-of-distribution (OOD) samples. In these situations, deep learning-based uncertainty metrics do not have to be indicative of the quality of the estimate\cite{kuleshov2018accurate} and methods that identify OOD situations should be employed to avoid trusting wrong estimations. In the field of multispectral imaging, OOD metrics were used to quantify domain gaps between data that a deep learning algorithm was trained on and newly acquired experimental data\cite{adler2019uncertainty,adler2019out}. We expect the investigation of well-calibrated methods for uncertainty estimation and the automatic detection of OOD scenarios to be integral towards the clinical translation of deep learning-based \ac{PAI} methodologies.\\

\textbf{Explainability.} The goal of the field of explainable deep learning is to provide a trace of the inference of developed algorithms \cite{holzinger2019causability}. The estimates of a deep learning algorithm should be comprehensible to domain experts in order to verify, improve, and learn from the system \cite{samek2017explainable}. In combination with techniques for uncertainty estimation and OOD detection, we believe that the explainability of algorithms will play an important role in the future of deep learning-based algorithms for PAI, especially in the biomedical context.\\

\textbf{Validation.} Thorough validation of methods is an integral part of clinical translation and as such plays a crucial role in the regulatory processes of medical device certification\cite{wilson2018challenges}. To this end, algorithms can be validated on several different levels, including in-distribution and out-of-distribution test data, as well as clinical validation in large-scale prospective studies\cite{steiner2018impact}. However, there is a systematic lack of prospective studies in the field of medical imaging with deep learning\cite{liu2019comparison}, and to our knowledge, there have been no such prospective deep learning studies in the field of \ac{PAI} yet. Some of the most impressive clinical trials in the field to date include the detection of Duchenne muscular dystrophy\cite{regensburger2019detection} and the assessment of disease activity in Crohn's disease\cite{knieling2017multispectral}. At least half of the reviewed papers have validated their methods on experimental data, but only approx. 20\% of papers have validated their methods on \emph{in vivo} data and even less on human measurements. This further demonstrates the vast differences in complexity within data obtained from phantoms versus living organisms. We expect that before deep learning methods for \ac{PAI} can reliably be used in a clinical context, much more pre-clinical work is needed to mature the proposed methodologies.\\

Another crucial aspect that we noticed during this review is the difficulty to compare many of the reported results. This is partly due to the fact that no standardized metrics or common data sets have so far been established in the field. Furthermore, the developed algorithms are tested only on in-house data sets that are usually not openly accessible. We have high hopes that these problems can be mitigated to a certain extent by the ongoing standardization efforts of the PA community, as promoted by the International Photoacoustic Standardisation Consortium (IPASC)\cite{bohndiek2019addressing}. Amongst other issues, this consortium is working on standardized methods to assess image quality and characterize \ac{PAI} device performance, on the organization of one of the first multi-centric studies in which PA phantoms are imaged all across the globe, as well as a standardized data format that facilitates the vendor-independent exchange of PA data.\\

\textbf{Computational efficiency.} Depending on the clinical use case, time can be of the essence (with stroke diagnosis being a prominent example\cite{fisher1995penumbra}) and the speed of the algorithm can be considered an important factor. \ac{PAI} is capable of real-time imaging\cite{gamelin2009real,kim2016programmable,kirchner2018signed} and the inference of estimates with deep learning can be exceptionally fast due to the massive parallelization capabilities of modern GPUs. The combination of these two factors can enable the real-time application of complex algorithms to \ac{PAI}. In the reviewed literature, it was demonstrated that entire high-resolution \ac{2D} and \ac{3D} images can be evaluated in a matter of milliseconds\cite{shang_two-step-training_2020}. In comparison to model-based methods, deep learning-based methods take a long time to train and fully optimize before they are ready to use. We believe that the drastic run-time performance increase could enable many time-critical applications of \ac{PAI} that might otherwise remain unfeasible.\\

\textbf{Clinical workflow integration.} Deep learning methods have already found success in several medical applications, especially in the field of radiology\cite{steiner2018impact,akkus2019survey,tonekaboni2019clinicians}. Nevertheless, we believe that the integrability of deep learning methods in \ac{PAI} heavily depends on the target clinical use case. The deep learning algorithm needs to have a clear impact on clinical practice, for example in terms of benefits for patients, personnel, or the hospital. Furthermore, the methods need to be easy to use for healthcare professionals, ideally being intuitive and introducing no significant time-burdens. \ac{PAI} is very promising for a multitude of clinical applications\cite{attia2019review}, which are mostly based on the differences in contrast based on local blood perfusion and blood oxygen saturation. To unleash to the full potential of \ac{PAI}, the inverse problems need to be solved to gain quantitative information on the underlying optical tissue properties. Deep learning can potentially enable an accurate, reliable, uncertainty-aware, and explainable estimation of the biomarkers of interest from the acquired PA measurements and thus provide unique opportunities towards the clinical translation of PAI. Nevertheless, thorough validation of the developed methods constitutes an essential first step in this direction.\\




\subsection{Conclusion}
This review has shown that deep learning methods possess unique advantages when applied to the field of \ac{PAI} and have the potential to facilitate its clinical translation in the long term. We analyzed the current state of the art of deep learning applications as pertaining to several open challenges in photoacoustic imaging: the acoustic and optical inverse problem, image post-processing, and semantic image annotation.\\

\textbf{Summary of findings:}
\begin{itemize}
    \item Deep learning methods in \ac{PAI} are currently still in their infancy. While the initial results are promising and encouraging, prospective clinical validation studies of such techniques, an integral part of method validation, have not been conducted.
    \item One of the core bottlenecks of the application of deep learning algorithms to \ac{PAI} is the lack of reliable, high-quality experimental training data. For this reason, about 75\% of deep learning papers in \ac{PAI} rely on simulated data for supervised algorithm training.
    \item A commonly used workaround to create suitable experimental training data for image post-processing is to artificially introduce artifacts, for example, by deliberately using less information for image reconstruction.
    \item Because the underlying optical tissue properties are inherently difficult to measure \emph{in vivo}, data-driven approaches towards the optical inverse problem have primarily relied on the presence of high-fidelity simulated data and have not yet successfully been applied \emph{in vivo}.
    \item While direct image reconstruction with deep learning shows exceptional promise due to the drastic speed increases compared to model-reconstruction schemes, deep learning methods that utilize additional information such as reconstructions from reference methods or hand-crafted feature vectors have proven much more generalizable.
    \item Approximately 50\% of papers test the presented methods on simulated data only and do not use multiple test sets that are significantly different from the training data distribution.
    \item A successful application of oxygenation estimation methods using entire \ac{2D} or \ac{3D} images has not yet been shown \emph{in vitro} or \emph{in vivo}. This is most probably caused by the large domain gap between synthetic and experimental PA images.
    \item Deep learning in \ac{PAI} has considerable room for improvement, for instance in terms of, generalizability, uncertainty estimation, out-of-distribution detection, or explainability.
\end{itemize}


\bibliography{literature}   
\bibliographystyle{spiejour}   


\section*{Acknowledgements}

This project has received funding from the European Union’s Horizon 2020 research and innovation programme through the ERC starting grant {\footnotesize COMBIOSCOPY} under grant agreement No. ERC-2015-StG-37960. The authors would like to thank M. D. Tizabi and A. Seitel for proof-reading the manuscript.

\section*{Author contributions statement}

Conceptualization, J.G., M.S., K.D., and L.M.-H.; Funding acquisition, L.M.-H.; Investigation, J.G., M.S., K.D., Methodology, J.G.; Project administration, J.G. and L.M.-H.; Writing - original draft, J.G.; and Writing - review and editing, J.G., M.S., K.D., and L.M.-H.


\end{document}